\documentclass[11pt]{article}

\usepackage{amsfonts,amssymb,chet,dsfont,graphicx,mathrsfs,pgfplots,tikz}

\usetikzlibrary{external}
\title{X-Ray Polarization Signals from Magnetars\\with Axion-Like-Particles}

\author{Jean-Fran\c{c}ois Fortin$^{\ast,}$\email{jean-francois.fortin@phy.ulaval.ca} and Kuver Sinha$^{\dagger,}$\email{kuver.sinha@ou.edu}}

\affiliation{
$^\ast$D\'epartement de Physique, de G\'enie Physique et d'Optique,\\Universit\'e Laval, Qu\'ebec, QC G1V 0A6, Canada\\
$^\dagger$Department of Physics and Astronomy, University of Oklahoma, Norman, OK 73019, USA
}

\abstract{%
Axion-like-particles (ALPs) produced in the core of a magnetar can convert to photons in the magnetosphere, giving rise to novel features in the X-ray spectrum.  Since ALPs only mix with the parallel mode of the photon, the polarization of the soft and hard X-ray spectra is predicted to have an O-mode component, in addition to the mainly X-mode component given by most astrophysical models.  The relative strength of the O-mode component depends on the intensity of ALPs produced in the core and the probability of conversion.  We quantify our results by considering X-ray emission produced both by astrophysical processes and by ALP-photon conversion, in an uncorrelated fashion, and in different relative proportions, which we parametrize by the angle $\chi_0$.  We then define a normalized astrophysics-subtracted Stokes parameter $R$ which only acquires non-zero values in the presence of ALP-photon conversion.  We find, remarkably, that the parameter $R$ factorizes into a product of the ALP-to-photon conversion probability and $\cos(2\chi_0)$ and display $R$, as well as the usual Stokes parameter $Q$, as a function of the photon energy and relative fractions of ALP and photon intensities.  For benchmark points currently allowed by the CAST experiment, the O-mode prediction can be tested in future X-ray polarimeters and used either to constrain ALPs or find evidence for them.
}

\date{July 2018} 

\begin{document}

\maketitle

\toc


\section{Introduction}\label{SIntro}

X-ray polarimetry is a nascent field that can explore different astrophysical sources, ranging from compact objects to pulsar wind nebulas, supernova remnants, and molecular clouds.  For a recent review of the astrophysical processes that can lead to polarized X-rays, as well as a description of current and planned missions, we refer to \cite{fabiani}.

The emission from compact objects, which are the focus of this article, can exhibit polarization due to the different opacities of the surface plasma to different components of the photon electric field.  However, polarization of X-ray radiation from neutron stars can also arise due to fundamental physics whose origin is distinct from astrophysics, and it is this possibility that we will entertain in our work.  It is worthwhile to remember that polarization experiments probe physical anisotropies, and there is an important anisotropy in the photon Lagrangian if axion-like-particles (ALPs) \cite{Peccei:1977hh,Peccei:1977ur,Wilczek:1977pj,Weinberg:1977ma} exist.  Namely, ALPs mix only with the parallel and not the perpendicular component of the electric field in the presence of an external magnetic field.  This anisotropy of the photon-ALP Lagrangian can thus be probed by X-ray polarimetry if the magnetic field near a compact object is strong enough.

Magnetars are an interesting subclass of neutron stars characterized by extremely strong magnetic fields, generally exceeding the quantum critical value $B_c=m_e^2/e=4.414\times10^{13}\,\text{G}$ \cite{Turolla:2015mwa,Beloborodov:2016mmx,Kaspi:2017fwg}, and constitute the natural target for our investigations.

In a previous paper \cite{Fortin:2018ehg}, we considered the production of ALPs from the core of magnetars and their subsequent conversion into photons (we refer to \cite{Lai:2006af,Chelouche:2008ta,Jimenez:2011pg,Perna:2012wn} for previous work in this direction).  The relevant terms in the ALP-photon Lagrangian are
\eqn{\mathcal{L}\supset-\frac{g}{4}aF_{\mu\nu}\tilde{F}^{\mu\nu}+g_{aN}(\partial_{\mu}a)\bar{N}\gamma^{\mu}\gamma_5N,}[EqnL]
where $a$ denotes the ALP and the coupling constants $g\equiv g_{a\gamma}$ and $g_{aN}$ have mass dimension $-1$.  The first term is responsible for ALP-photon conversion in an external magnetic field \cite{Raffelt:1987im,Marsh:2015xka,Graham:2015ouw}, while the second term in \eqref{EqnL} is the coupling between the ALP and nucleons $N$ that leads to ALP production in the core of neutron stars.  As mentioned before, ALPs mix with the component of the electric field that lies in the plane containing the external magnetic field $B$ and the radial direction of motion, while the perpendicular component of the electric field propagates unaffected.  This implies that the polarization pattern of the observed spectrum will be affected by the presence of ALPs.

The purpose of this paper is to investigate the changes in polarization patterns in the observed X-ray spectra of magnetars in the presence of ALPs.  Our approach consists of the following steps.  Firstly, we assume that for every frequency $\omega\sim1-\mathcal{O}(\text{few hundred})\,\text{keV}$, both photons and ALPs are produced near the surface of the magnetar, in an uncorrelated manner.  The important observation is that the astrophysical processes leading to photon production both in the soft as well as the hard X-ray regime are completely independent of the processes that give rise to ALPs, mainly by nucleon-nucleon bremsstrahlung \cite{Iwamoto:1984ir,Iwamoto:1992jp,Nakagawa:1987pga,Nakagawa:1988rhp,Raffelt:1996wa,Umeda:1997da,Paul:2018msp,Maruyama:2017xzl}.

The evolution of the ALP-photon system in the magnetic field of the magnetar is conveniently described in terms of a set of first-order differential equations coupling the amplitudes and phase difference of the ALP and photon fields.  This parametrization has been shown to drastically simplify numerical analyses of the system \cite{Fortin:2018ehg}.  The uncorrelated production mechanism of ALPs and photons alluded to above allows us to average over the initial phase differences.  On the other hand, we parametrize the initial relative amplitudes by the angle $\chi_0$.  For example, $\chi_0=0$ signifies a pure ALP initial state.  The introduction of $\chi_0$ allows us to remain agnostic about astrophysical models of production of X-ray photons near a magnetar's surface.

Our second step is to define an astrophysics-subtracted (or surface-subtracted) normalized Stokes parameter $R$.  This is done as follows.  The sum ($I=\bar{I}_\perp+\bar{I}_\parallel$) and difference ($Q=\bar{I}_\perp-\bar{I}_\parallel$) of the phase-averaged photon intensities in the parallel and perpendicular planes are first computed, and then the quantities $\Delta I$ and $\Delta Q$ are defined, which are the values of the respective Stokes parameters away from the magnetar, minus their values at the magnetar's surface $r_0$.  This is very useful, since at the surface (which we take to also include the plasma) astrophysical effects due to differential opacities of the plasma lead to polarization, which should be subtracted away to extract the contribution to the polarization coming purely from ALP-photon mixing in the magnetosphere.  In the absence of ALPs (\textit{i.e.} if ALPs do not exist in the Universe), and assuming that astrophysical processes in the magnetosphere do not lead to substantial polarization, the surface-subtracted Stokes parameters should vanish.

We find, remarkably, that the surface-subtracted Stokes parameters $\Delta I$ and $\Delta Q$ factorize into two pieces: the ALP-to-photon conversion probability and a factor that encodes the composition of the initial state.  Specfically, normalizing the surface-subtracted Stokes parameter and defining the quantity
\eqn{R=\frac{\Delta I}{A^2},}
where $A$ is related to the initial amplitudes of the ALP and parallel photon states, we find that
\eqn{R=P_{a\to\gamma}\cos(2\chi_0).}[EqnRfact]

This is our main result.  It has several implications.  Firstly, the spatial dependence of the Stokes parameters, as well as their dependence on the properties of the magnetar and the ALP parameters like the mass and coupling constant, are all encoded in the ALP-to-photon conversion probability $P_{a\to\gamma}$.  Secondly, given the conversion probability, one can obtain the value of $R$ simply by scaling with the appropriate initial condition $\cos(2\chi_0)$.  Most importantly, the factorization of $R$ in \eqref{EqnRfact} can be utilized to provide an analytic expression for the Stokes parameter $Q$ in \eqref{EqnStokesSoln}, which is the parameter we prefer when discussing the observational aspects of our work.

We utilize the methods of our previous paper \cite{Fortin:2018ehg} to display the dependence of $R$ on the photon energy as well as ALP parameters.  As a benchmark point, we take the CAST-allowed values for the ALP mass and coupling, $m_a=10^{-8}\,\text{keV}$ and $g/e=5\times10^{-17}\,\text{keV}^{-1}$, respectively.  The dependence of $Q$ on $\bar{I}_\parallel/\bar{I}_\perp$ and $\bar{I}_a/\bar{I}_\perp$ is shown in Fig.~\ref{FigStokesQ}, where $\bar{I}_a$ is the phase-averaged ALP intensity.

In the absence of ALPs, astrophysical modeling of thermal and hard X-rays from magnetars predicts mainly X-mode polarization, for which the electric field is perpendicular to the plane containing the external magnetic field and the direction of propagation.  For the strong magnetic fields of magnetars, the polarization in the X-mode is expected to be especially pronounced due to the vacuum birefringence effect.  The polarization radius, which is the location where the polarization vector stops tracking the magnetic field, is large for strong magnetic fields and the overall polarization is enhanced.

The unique observational signature of ALP-photon conversion is the change in the predicted polarization pattern.  ALPs add to the astrophysical picture described above by producing O-mode photons, for which the electric field is parallel to the plane containing the external magnetic field and the direction of propagation.  We compute the radius of conversion, where the probability of conversion becomes significant, and find that it is typically of the same order or larger than the polarization radius, implying an overall O-mode superposed on the X-mode coming purely from astrophysics.  These results are displayed in terms of the Stokes parameter $Q$ in Fig.~\ref{FigQ}.  The astrophysical prediction of $Q$ is expected to be $Q\sim\bar{I}_\perp$, while the presence of ALP-to-photon conversion drives $Q$ to be smaller, and perhaps even negative depending on the intensity of ALPs produced from the core.  The next generation of X-ray polarimeters \cite{polstar,ixpe} and increasingly sophisticated modeling of the astrophysics of magnetars \cite{Lai:2006af,Chelouche:2008ta,Jimenez:2011pg,Perna:2012wn,Wadiasingh:2017rcq} provide an opportunity to investigate ALPs using polarization.

Our paper is organized as follows.  In Section \ref{SOsc}, we recapitulate the ALP-photon coupled system and the results of \cite{Fortin:2018ehg}.  In Section \ref{SPP}, we perform the calculation of the Stokes parameters.  In Section \ref{SResults}, we find analytical approximations for the parameter $R$ and display it as a function of the photon energy and ALP-photon coupling.  We also describe the observational possibilities and contrast our results with previous work in Section \ref{SObs}.  We end with our conclusions in Section \ref{SConclusion}.  Finally, Appendix \ref{SApp} demonstrates the factorization property of the normalized surface-subtracted Stokes parameter $R$.
 

\section{Oscillations}\label{SOsc}

In this section we introduce the evolution equations for the photon and the ALP relevant to magnetars.  We use the general formalism of \cite{Fortin:2018ehg} for oscillations in the limits where the space variations of the magnetic field are large compared to the particle wavelength and dispersion is weak.  The reader interested in the details of this formalism is referred to \cite{Fortin:2018ehg}.

\subsection{Evolution Equations}

For particles with energies in the $1$ to $200$ keV range (corresponding to soft and hard X-ray photons) propagating radially outwards from a magnetar, the system is in the appropriate limits (as long as the magnetic field is not too large) with negligible plasma contributions to the evolution equations \cite{Raffelt:1987im,Lai:2006af}, leading to
\eqn{i\frac{d}{dx}\left(\begin{array}{c}a\\E_\parallel\\E_\perp\end{array}\right)=\left(\begin{array}{ccc}\omega r_0+\Delta_ar_0&\Delta_Mr_0&0\\\Delta_Mr_0&\omega r_0+\Delta_\parallel r_0&0\\0&0&\omega r_0+\Delta_\perp r_0\end{array}\right)\left(\begin{array}{c}a\\E_\parallel\\E_\perp\end{array}\right).}[EqnDiffMat]
where
\eqn{\Delta_a=-\frac{m_a^2}{2\omega},\qquad\qquad\Delta_\parallel=\frac{1}{2}q_\parallel\omega\sin^2\theta,\qquad\qquad\Delta_\perp=\frac{1}{2}q_\perp\omega\sin^2\theta,\qquad\qquad\Delta_M=\frac{1}{2}gB\sin\theta.}
Here $a(x)$, $E_\parallel(x)$ and $E_\perp(x)$ are the ALP and parallel and perpendicular photon electric fields respectively while $x=r/r_0$ with $r$ the distance from the center of the magnetar and $r_0$ the magnetar's radius.  Moreover, $\omega$ is the energy of the particles, $m_a$ is the ALP mass, $g$ is the ALP-photon coupling constant and $\theta$ is the angle between the magnetic field and the direction of propagation of the particle.  $q_\parallel$ and $q_\perp$ are dimensionless functions of the magnetic field $B$ given by \cite{Lai:2006af,Raffelt:1987im}
\eqn{
\begin{gathered}
q_\parallel=\frac{7\alpha}{45\pi}b^2\hat{q}_\parallel,\qquad\qquad\hat{q}_\parallel=\frac{1+1.2b}{1+1.33b+0.56b^2},\\
q_\perp=\frac{4\alpha}{45\pi}b^2\hat{q}_\perp,\qquad\qquad\hat{q}_\perp=\frac{1}{1+0.72b^{5/4}+(4/15)b^2},
\end{gathered}
}
with $b=B/B_c$ where $B_c=m_e^2/e=4.414\times10^{13}\,\text{G}$ is the critical QED field strength.  Here $e=\sqrt{4\pi\alpha}$ where the fine structure constant is approximatively $\alpha\approx1/137$.

Since the plasma contributions are negligible, the three-state system \eqref{EqnDiffMat} effectively decomposes into two independent oscillation systems, a two-state system for the ALP and the parallel photon, and a one-state system for the perpendicular photon.

Due to the probability conservation property $\frac{d}{dx}[|a(x)|^2+|E_\parallel(x)|^2]=0$ discussed in \cite{Fortin:2018ehg}, the different states can be expressed as
\eqn{a(x)=A\cos[\chi(x)]e^{-i\phi_a(x)},\qquad\qquad E_\parallel(x)=iA\sin[\chi(x)]e^{-i\phi_\parallel(x)},\qquad\qquad E_\perp(x)=A_\perp e^{-i\phi_\perp(x)},}[EqnaEE]
where $A_a=A\cos[\chi(x)]$, $A_\parallel=A\sin[\chi(x)]$ and $A_\perp$ are the amplitudes at position $xr_0$ of the ALP field, the parallel photon field and the perpendicular photon field respectively.  It is important to note that $A$ and $A_\perp$ are constants which can always be chosen real and positive while $\chi(x)$, $\phi_a(x)$, $\phi_\parallel(x)$ and $\phi_\perp(x)$ are real functions.  Hence, the intensities at position $xr_0$ are given by the respective amplitudes squared, \textit{i.e.} $I_a(x)=A^2\cos^2[\chi(x)]$, $I_\parallel(x)=A^2\sin^2[\chi(x)]$ and $I_\perp(x)=A_\perp^2$ for the ALP field, the parallel photon field and the perpendicular photon field.  Probability conservation then implies that $I_a(x)+I_\parallel(x)=A^2$ and $I_\perp(x)=A_\perp^2$ are constants.

Using \eqref{EqnaEE} in \eqref{EqnDiffMat}, the evolution equations become
\eqna{
\frac{d\chi(x)}{dx}&=-\Delta_Mr_0\cos[\Delta\phi(x)],\\
\frac{d\Delta\phi(x)}{dx}&=(\Delta_a-\Delta_\parallel)r_0+2\Delta_Mr_0\cot[2\chi(x)]\sin[\Delta\phi(x)],\\
\frac{d\Sigma\phi(x)}{dx}&=(2\omega+\Delta_a+\Delta_\parallel)r_0-2\Delta_Mr_0\csc[2\chi(x)]\sin[\Delta\phi(x)],\\
\frac{d\phi_\perp(x)}{dx}&=(\omega+\Delta_\perp)r_0,
}[EqnEE]
where $\Delta\phi(x)=\phi_a(x)-\phi_\parallel(x)$ is the phase difference between the ALP field and the parallel photon field while $\Sigma\phi(x)=\phi_a(x)+\phi_\parallel(x)$ is the sum of the phases of the ALP field and the parallel photon field.  In addition to showing that the differential equation for $\phi_\perp(x)$ decouples, the evolution equations \eqref{EqnEE} imply also that the differential equation for $\Sigma\phi(x)$ decouples in the sense that it is completely determined once the solutions to the coupled $\chi(x)$ and $\Delta\phi(x)$ differential equations are known.  Hence both $\Sigma\phi(x)$ and $\phi_\perp(x)$ are irrelevant in computing the intensities.

The initial states are determined at the surface of the magnetar from the boundary conditions at $x=1$.  For the two-state system described by $\chi(x)$ and $\Delta\phi(x)$, pure initial states satisfy $\chi(1)=n\pi/2$ with $n\in\mathbb{Z}$, and the boundary condition for $\Delta\phi(1)$ must satisfy $\Delta\phi(1)=m\pi$ with $m\in\mathbb{Z}$ to avoid singularities.  The two different choices of phase difference for pure initial states lead to the same intensities.

Indeed, it is straightforward to verify that the transformation
\eqn{\chi(x)\to-\chi(x),\qquad\qquad\Delta\phi(x)\to\Delta\phi(x)\pm\pi,}
leaves the evolution equations \eqref{EqnEE} invariant.  Therefore, for boundary conditions given by $\chi(1)=\chi_0$ and $\Delta\phi(1)=\Delta\phi_0$, the intensities verify
\eqn{I_a(\chi_0,\Delta\phi_0,x)=I_a(-\chi_0,\Delta\phi_0\pm\pi,x),\qquad\qquad I_\parallel(\chi_0,\Delta\phi_0,x)=I_\parallel(-\chi_0,\Delta\phi_0\pm\pi,x).}
Thus, for the pure ALP initial state $\chi_0=0$, the two different choices of phase difference, say $\Delta\phi_0=0$ and $\Delta\phi_0=\pi$, give the same intensities.  For the pure parallel photon initial state $\chi_0=\pi/2$, the two different choices of phase difference, say $\Delta\phi_0=0$ and $\Delta\phi_0=\pi$ again, give the same intensities since \eqref{EqnEE} are also invariant under the transformation $\chi(x)\to\chi(x)\pm\pi$.  By using such arguments, more can be said about the intensities.

Indeed, two other transformations play an important role in the following.  First, the transformation
\eqn{\chi(x)\to\pi-\chi(x),\qquad\qquad\Delta\phi(x)\to\Delta\phi(x)\pm\pi,}[EqnTpi]
leaves the evolution equations \eqref{EqnEE} invariant, which implies that the intensities satisfy the following relations,
\eqn{I_a(\chi_0,\Delta\phi_0,x)=I_a(\pi-\chi_0,\Delta\phi_0\pm\pi,x),\qquad\qquad I_\parallel(\chi_0,\Delta\phi_0,x)=I_\parallel(\pi-\chi_0,\Delta\phi_0\pm\pi,x).}[EqnTpiI]

Second, in general in the definitions \eqref{EqnaEE}, one can always take $\chi(x)\in[0,\pi/2]$.  With that interval in mind, it is interesting to note that $\pi/2-\chi(x)\in[0,\pi/2]$.  Hence, applying the transformation
\eqn{\chi(x)\to\pi/2-\chi(x),\qquad\qquad\Delta\phi(x)\to\Delta\phi(x)\pm\pi,}[EqnTpi2]
leaves the evolution equations of the coupled $\chi(x)$ and $\Delta\phi(x)$ invariant while keeping the boundary condition on $\chi(x)$ in the appropriate interval.  Therefore, it is easy to conclude that if $\chi(x)$ and $\Delta\phi(x)$ are solutions to \eqref{EqnEE}, then from \eqref{EqnTpi2} $\pi/2-\chi(x)$ and $\Delta\phi(x)\pm\pi$ are also solutions to \eqref{EqnEE}.  As before, this implies some conditions on the intensities which are here given by
\eqna{
I_a(\chi_0,\Delta\phi_0,x)&=I_\parallel(\pi/2-\chi_0,\Delta\phi_0\pm\pi,x)=A^2-I_a(\pi/2-\chi_0,\Delta\phi_0\pm\pi,x),\\
I_\parallel(\chi_0,\Delta\phi_0,x)&=I_a(\pi/2-\chi_0,\Delta\phi_0\pm\pi,x)=A^2-I_\parallel(\pi/2-\chi_0,\Delta\phi_0\pm\pi,x).
}[EqnTpi2I]
It is important to note that \eqref{EqnTpi2I} are true simply because the intensities do not care about the sum of the phases.  Indeed, if $\Sigma\phi(x)$ had been important for the intensities, \textit{i.e.} if it had not decoupled, then the relations \eqref{EqnTpi2I} would not be true because the transformation \eqref{EqnTpi2} does not leave the differential equation for $\Sigma\phi(x)$ in \eqref{EqnEE} invariant.

\subsection{Mixed Initial States}

As discussed in the introduction, in the X-ray regime, production mechanisms in magnetars have very different origins for ALPs and photons.  The former comes predominantly from nucleon-nucleon bremsstrahlung of ALPs \cite{Iwamoto:1984ir,Iwamoto:1992jp,Nakagawa:1987pga,Nakagawa:1988rhp,Raffelt:1996wa,Umeda:1997da,Paul:2018msp,Maruyama:2017xzl} while the latter originates from the magnetar itself and interactions with the plasma in the magnetar's atmosphere (\textit{e.g.} parallel-perpendicular mode conversion in inhomogeneous magnetar's atmosphere) \cite{Lai:2006af}.

Hence, it is expected that at the magnetar's surface, where the ALP and parallel photon fields start oscillating into one another as they travel outwards, their amplitudes can differ substantially depending on the actual values of the ALP parameters.  Moreover, their phase difference is effectively random.

Thus, for given initial amplitudes corresponding to specific ALP parameters, \textit{i.e.} for a specific $A$ and $\chi(1)=\chi_0$, it is natural to average over the initial phase difference $\Delta\phi(1)=\Delta\phi_0$ to determine the intensities away from the magnetar.  This implies that, at a distance $xr_0$ from the magnetar's surface, the averaged intensities $\bar{I}(x)$ for initial $\chi(1)=\chi_0$ are
\eqn{\bar{I}_a(\chi_0,x)=\int_0^{2\pi}\frac{d\Delta\phi_0}{2\pi}\,I_a(\chi_0,\Delta\phi_0,x),\qquad\qquad\bar{I}_\parallel(\chi_0,x)=\int_0^{2\pi}\frac{d\Delta\phi_0}{2\pi}\,I_\parallel(\chi_0,\Delta\phi_0,x).}[EqnIb]
Due to the singularities appearing for pure initial states and the equality of the intensities for the two different choices of phase difference, there is no averaging for pure initial states and only one initial phase difference (say $\Delta\phi_0=0$) is necessary to determine the averaged intensities at position $xr_0$.

Since the intensities transform as in \eqref{EqnTpiI} under the transformation \eqref{EqnTpi} and \eqref{EqnTpi2I} under the transformation \eqref{EqnTpi2}, the averaged intensities \eqref{EqnIb} also behave as
\eqn{\bar{I}_a(\chi_0,x)=\bar{I}_a(\pi-\chi_0,x),\qquad\qquad\bar{I}_\parallel(\chi_0,x)=\bar{I}_\parallel(\pi-\chi_0,x).}[EqnTpiIb]
and
\eqna{
\bar{I}_a(\chi_0,x)&=\bar{I}_\parallel(\pi/2-\chi_0,x)=A^2-\bar{I}_a(\pi/2-\chi_0,x),\\
\bar{I}_\parallel(\chi_0,x)&=\bar{I}_a(\pi/2-\chi_0,x)=A^2-\bar{I}_\parallel(\pi/2-\chi_0,x).
}[EqnTpi2Ib]
Moreover, from the invariance under the transformation $\chi(x)\to\chi(x)\pm\pi$, the averaged intensities are periodic functions of $\chi_0$ with period $\pi$.  Hence from \eqref{EqnTpiIb} the averaged intensities are even functions of $\chi_0$ with respect to $\pi/2$ while from \eqref{EqnTpi2Ib} the averaged intensities for some initial amplitudes are related to the averaged intensities for the inverted initial amplitudes.  As a corollary of \eqref{EqnTpi2Ib}, it is straightforward to see that $\bar{I}_a(\pi/4,x)=\bar{I}_\parallel(\pi/4,x)=A^2/2$.  Therefore, for a half-and-half initial state, the averaged intensities are exactly $1/2$ of the sum of the ALP and parallel photon initial intensities.  Thus, it is only necessary to compute the averaged intensities for initial $\chi(1)\in[0,\pi/4)$ to determine the averaged intensities for all possible initial states.  In fact, the next section demonstrates that we can do better.

Obviously, averaging over the perpendicular photon initial phase is of no consequence for the averaged perpendicular photon intensity, hence $\bar{I}_\perp(x)=A_\perp^2$ and the averaged perpendicular photon intensity remains constant.


\section{Photon Polarizations}\label{SPP}

This section introduces the Stokes parameters describing the polarization state of the photon signal.  The effect of ALP-photon coupling is discussed qualitatively, with a more quantitative analysis for a particular magnetar presented in the next section.

\subsection{Stokes Parameters}

At a given position $xr_0$ from the magnetar, the relevant Stokes parameters are $I$ and $Q$ which correspond respectively to the sum of and the difference between the averaged perpendicular photon intensity and the averaged parallel photon intensity, \textit{i.e.}
\eqn{I(\chi_0,x)=\bar{I}_\perp(x)+\bar{I}_\parallel(\chi_0,x),\qquad\qquad Q(\chi_0,x)=\bar{I}_\perp(x)-\bar{I}_\parallel(\chi_0,x).}[EqnStokes]
As pointed out before, since the perpendicular photon field does not mix with the ALP-parallel photon two-state system, its averaged intensity is constant.  Therefore, all modifications to the Stokes parameter are driven by the averaged parallel photon intensity.

Moreover, since plasma contributions are negligible outside the magnetar's atmosphere, in the absence of ALPs the evolution equations \eqref{EqnDiffMat} correspond to two independent one-state systems, one for the parallel photon field and one for the perpendicular photon field.  Hence without ALPs the Stokes parameters at a distance $xr_0$ from the magnetar would be the same as the Stokes parameters at the magnetar's surface.  The effects of the existence of ALPs and the possible photon oscillations with them on the Stokes parameters can therefore be conveniently analysed by studying the differences between the Stokes parameters at $xr_0$ \eqref{EqnStokes} and the Stokes parameters at $r_0$, \textit{i.e.}
\eqna{
\Delta I(\chi_0,x)&=I(\chi_0,x)-I(\chi_0,1)=\bar{I}_\parallel(\chi_0,x)-A^2\sin^2(\chi_0),\\
\Delta Q(\chi_0,x)&=Q(\chi_0,x)-Q(\chi_0,1)=-[\bar{I}_\parallel(\chi_0,x)-A^2\sin^2(\chi_0)]=-\Delta I(\chi_0,x).
}[EqnStokesDiff]
The relation $\Delta Q(\chi_0,x)=-\Delta I(\chi_0,x)$ demonstrates perfect anti-correlation between the surface-subtracted Stokes parameter $\Delta I(\chi_0,x)$ and the surface-subtracted Stokes parameter $\Delta Q(\chi_0,x)$.  Therefore, in our scenario, there is only one independent quantity to keep track of, which we choose as the surface-subtracted Stokes parameter $\Delta I(\chi_0,x)$.

It is clear from this analysis and especially \eqref{EqnStokesDiff} that an astrophysical understanding of soft and hard X-ray polarized emission from magnetars can constrain or even discover ALPs.  Indeed, on the one hand standard astrophysical considerations dictate the expected Stokes parameters $I^\text{exp}$ and $Q^\text{exp}$ at the magnetar's surface.  On the other hand, observations on Earth lead to observed Stokes parameters $I^\text{obs}$ and $Q^\text{obs}$.  Thus theoretical and observational astrophysics determine the differences of the Stokes parameters $\Delta I^\text{astro}=I^\text{obs}-I^\text{exp}$ and $\Delta Q^\text{astro}=Q^\text{obs}-Q^\text{exp}$.  Non-vanishing values for $\Delta I^\text{astro}$ and $\Delta Q^\text{astro}$ suggest either a misunderstanding of the astrophysical processes at play in magnetars or observational errors.  However, from \eqref{EqnStokesDiff} non-vanishing values for $\Delta I^\text{astro}$ and $\Delta Q^\text{astro}$ such that $\Delta Q^\text{astro}=-\Delta I^\text{astro}$ strongly suggest that ALPs exist and ALP-photon oscillations occur in the magnetic field of the magnetar.  The relation $\Delta Q^\text{astro}=-\Delta I^\text{astro}$ originating from \eqref{EqnStokesDiff} can thus be seen as a smoking gun signal for ALPs.

Before turning to an analysis of the transformation properties of the surface-subtracted Stokes parameters and their consequences, it is of interest to define a new quantity, the normalized surface-subtracted Stokes parameter $R(\chi_0,x)$.

\subsection{Normalized Surface-Subtracted Stokes Parameter}

Since we do not rely on any specific models for the ALP and photon production mechanisms, the values of $A$, $\chi_0$ and $A_\perp$ which determine the initial ALP and photon amplitudes at the magnetar's surface are not fixed.  However, from \eqref{EqnStokesDiff} it is clear that the differences of the Stokes parameters at infinity and at the magnetar's surface does not depend on $A_\perp$.  Moreover, since $\bar{I}_\parallel(\chi_0,x)\propto A^2$, their dependence on $A$ is simple.  It is thus convenient in the following to study the normalized surface-subtracted Stokes parameter
\eqn{R(\chi_0,x)=\frac{\Delta I(\chi_0,x)}{A^2}=\cos(2\chi_0)\int_0^{2\pi}\frac{d\Delta\phi_0}{2\pi}\,\frac{1}{2}\left\{1-\frac{\cos\left[\left.2\chi(x)\right|_{\chi(1)=\chi_0,\Delta\phi(1)=\Delta\phi_0}\right]}{\cos(2\chi_0)}\right\}.}[EqnR]
Alternatively, one can set $A=1$ in \eqref{EqnStokesDiff} to compute the normalized surface-subtracted Stokes parameter \eqref{EqnR}.  Apart from the implicit dependence on the ALP and magnetar parameters, the normalized surface-subtracted Stokes parameter \eqref{EqnR} is thus a function of the boundary condition $\chi_0$ only instead of the boundary conditions $A$, $\chi_0$ and $A_\perp$.

The transformation properties of the normalized surface-subtracted Stokes parameter (and its non-normalized counterpart) are easily obtained.  They are
\eqna{
R(\chi_0,x)&=R(\chi_0\pm\pi,x),\\
R(\chi_0,x)&=R(\pi-\chi_0,x),\\
R(\chi_0,x)&=-R(\pi/2-\chi_0,x),
}[EqnTR]
from the invariance of \eqref{EqnEE} under the transformations $\chi(x)\to\chi(x)\pm\pi$, \eqref{EqnTpi} and \eqref{EqnTpi2} respectively.  Therefore \eqref{EqnTR} implies that the normalized surface-subtracted Stokes parameter is a periodic function of $\chi_0$ with period $\pi$ which is even in $\chi_0$ with respect to $\pi/2$ and odd in $\chi_0$ with respect to $\pi/4$.  As such, it can be expanded in Fourier modes as
\eqn{R(\chi_0,x)=\sum_{n\geq0}R_n(x)\cos[2(2n+1)\chi_0],}[EqnRF]
with the modes given by
\eqn{R_n(x)=\int_0^\pi\frac{d\chi_0}{\pi}\,2\cos[2(2n+1)\chi_0]R(\chi_0,x).}[EqnRFM]
The latter can be computed directly from \eqref{EqnR} since the integral is $\chi_0$-independent.  Indeed, averaging over $\Delta\phi_0$ eliminates all dependence on the boundary conditions (this statement is highly non-trivial, for a proof the reader is referred to Appendix \ref{SApp}).  Then the Fourier modes \eqref{EqnRFM} are simply
\eqn{R_n(x)=\delta_{n0}\int_0^{2\pi}\frac{d\Delta\phi_0}{2\pi}\,\frac{1}{2}\left\{1-\frac{\cos\left[\left.2\chi(x)\right|_{\chi(1)=\chi_0,\Delta\phi(1)=\Delta\phi_0}\right]}{\cos(2\chi_0)}\right\},}[EqnRFMSoln]
which are $\chi_0$-independent and thus the normalized surface-subtracted Stokes parameter \eqref{EqnRF} is
\eqn{R(\chi_0,x)=R_0(x)\cos(2\chi_0).}[EqnRFSoln]
It can be written in a more familiar form by setting $\chi_0=0$.  Indeed, since \eqref{EqnRFMSoln} is $\chi_0$-independent, we can set $\chi_0=0$ and replace the average over $\Delta\phi_0$ by $\Delta\phi_0=0$.  Analogously, we can set $\chi_0=0$ directly in \eqref{EqnRFSoln} and compare both sides.  Both approaches lead to our final result
\eqn{R(\chi_0,x)=R(\chi_0=0,x)\cos(2\chi_0)=\sin^2\left[\left.\chi(x)\right|_{\chi(1)=0,\Delta\phi(1)=0}\right]\cos(2\chi_0)=P_{a\to\gamma}(x)\cos(2\chi_0),}[EqnRSoln]
where
\eqn{P_{a\to\gamma}(x)=\sin^2\left[\left.\chi(x)\right|_{\chi(1)=0,\Delta\phi(1)=0}\right],}
is the ALP-to-photon conversion probability at a distance $xr_0$ for pure ALP initial state.  Therefore, the normalized surface-subtracted Stokes parameter factorizes into two parts, the ALP-to-photon conversion probability for pure ALP initial state $P_{a\to\gamma}(x)$ that is a function of the ALP and magnetar parameters, and $\cos(2\chi_0)$ which encodes the dependence on the mixed initial state.  In addition, from \eqref{EqnRSoln} it is clear that $-1\leq R(\chi_0,x)\leq1$ with $R(\pi/4,x)=0$ as proven before.

Using \eqref{EqnRSoln}, the surface-subtracted Stokes parameters \eqref{EqnStokesDiff} are
\eqn{\Delta I(\chi_0,x)=-\Delta Q(\chi_0,x)=A^2R(\chi_0,x)=A^2P_{a\to\gamma}(x)\cos(2\chi_0),}[EqnStokesDiffSoln]
which implies that the Stokes parameters \eqref{EqnStokes} and the averaged intensities \eqref{EqnIb} are
\eqna{
I(\chi_0,x)&=A_\perp^2+\frac{A^2}{2}\left\{1+[2P_{a\to\gamma}(x)-1]\cos(2\chi_0)\right\},\\
Q(\chi_0,x)&=A_\perp^2-\frac{A^2}{2}\left\{1+[2P_{a\to\gamma}(x)-1]\cos(2\chi_0)\right\},
}[EqnStokesSoln]
and
\eqna{
\bar{I}_a(\chi_0,x)&=\frac{A^2}{2}\left\{1-[2P_{a\to\gamma}(x)-1]\cos(2\chi_0)\right\},\\
\bar{I}_\parallel(\chi_0,x)&=\frac{A^2}{2}\left\{1+[2P_{a\to\gamma}(x)-1]\cos(2\chi_0)\right\},\\
\bar{I}_\perp(\chi_0,x)&=A_\perp^2,
}[EqnIbSoln]
respectively.

The behavior on the mixture of the initial state has been completely uncovered in \eqref{EqnRSoln}, \eqref{EqnStokesDiffSoln}, \eqref{EqnIbSoln} and \eqref{EqnStokesSoln}.  Indeed, the dependences on $A$, $\chi_0$ and $A_\perp$ are exhibited explicitly.  Hence it is easy to determine the behavior of the averaged intensities or the Stokes parameters under changes of the mixed initial state.  For example, the Stokes parameter $Q(\chi_0,x)$ at a distance $xr_0$ is always smaller (larger) than or equal to the Stokes parameter $Q(\chi_0,1)$ at the magnetar's surface if $\chi_0\in[0,\pi/4]$ ($\chi_0\in[\pi/4,\pi/2]$), \textit{i.e.} if the initial state is a mixture dominated by ALP (parallel photon).

Moreover, the ALP-to-photon conversion probability for pure ALP initial state is the only necessary quantity to compute in order to determine all the remaining quantities of interest.  This important observation simplifies greatly the problem since $P_{a\to\gamma}(x)$ is independent of $\chi_0$ and no averaging on $\Delta\phi_0$ is necessary.

The remarkable factorization of the dependence on the initial state in the averaged intensities is a consequence of the averaging over the initial phase difference.  If the average had not been performed, there would have been no such factorization.  Hence, if there exists a common origin for the production of ALPs and photons, then the initial phase difference would be fixed and the dependence on $\chi_0$ would not be as straightforward as in \eqref{EqnIbSoln}.

Finally, since the argument leading to the results above relies only on the probability conservation property of the system, it is important to note that relations analogous to \eqref{EqnIbSoln} for averaged intensities exist for any two-state oscillation system satisfying the probability conservation property.  Although we have not investigated it further, we expect that relations analogous to \eqref{EqnIbSoln} for averaged intensities can be found for any $n$-state oscillation system as long as probability is conserved.


\section{Results}\label{SResults}

In this section we solve numerically the evolution equations \eqref{EqnEE} with dipole magnetic field (valid a few $r_0$ away from the magnetar) for magnetar parameters corresponding to SGR 1806-20 in the soft and hard X-ray range.  As mentioned in the introduction, we stay agnostic with respect to the different production mechanisms for ALPs and photons.  Our results are therefore conveniently expressed with the help of the normalized surface-subtracted Stokes parameter \eqref{EqnR} far away from the magnetar where the magnetic field is effectively turned off.

In principle, \eqref{EqnR} forces us to compute $R$ by averaging over the initial phase difference for several mixed initial states.  However, the relation \eqref{EqnRSoln} allows us to focus only on the ALP-to-photon conversion probability for pure ALP initial state.  Nevertheless, to verify the relation \eqref{EqnRSoln}, we compute (for various ALP and magnetar parameters) $R$ the hard way using \eqref{EqnR} and compare with the ALP-to-photon conversion probability for pure ALP initial state through \eqref{EqnRSoln}, validating the argument of the previous section.

For SGR 1806-20, the ALP-to-photon conversion probability for pure ALP initial state in the soft and hard X-ray range was studied in \cite{Fortin:2018ehg}.  For completeness, we include here the relevant plots as well as a discussion of the relevant features of $R(\chi_0=0)\equiv R(\chi_0=0,x=\infty)$ for a benchmark point that is not excluded by CAST \cite{Anastassopoulos:2017ftl}.

\subsection{Dependence on the ALP and Magnetar Parameters}

As mentioned previously, standard astrophysical considerations dictates how the magnetar and its atmosphere behave.  Hence, theoretical astrophysics determines the initial amplitudes of the particles at the magnetar's surface.  The particles then travel through the plasma in the magnetar's atmosphere, which is of the order of centimeters, where they can encounter a resonance in the parallel-perpendicular photon space \cite{Lai:2006af}.  Since we remain agnostic about the astrophysical processes at play, we can start evolving the evolution equations \eqref{EqnEE} once the particles exit the magnetar's atmosphere, which is the only region where plasma effects are non-negligible.  Therefore, our initial amplitudes at the magnetar's surface correspond more precisely to the amplitudes outside the magnetar's atmosphere.  However, since the magnetar radius is of the order of kilometers while the atmosphere's thickness is of the order of centimeters, we can choose $x=1$ as our initial point to solve the evolution equations \eqref{EqnEE} without consequences.  From this point on, the plasma is completely negligible and the analysis presented before is valid.

However, the dipolar approximation holds only several radii away from the magnetar.  Hence, to assume a dipole magnetic field, it is necessary to redefine the initial amplitudes at the magnetar's surface by the amplitudes several radii away from the magnetar where the dipolar approximation is valid.  The evolution equations \eqref{EqnEE} are thus evolved numerically from there and the average in \eqref{EqnR} is performed on the phase difference at that position.  Since the conversion radius, where most of the ALP-parallel photon conversion occurs, is a few hundred radii away from the magnetar for our benchmark point, the dipolar approximation is clearly justified.

The particles thus evolve through an astrophysical sector consisting of the production region (either inside or closely by the magnetar), the plasma region and the region where the dipolar approximation is not valid.  The amplitudes at the end of this sector are then used as initial amplitudes for the particles evolving through a vacuum sector consisting of the dipolar magnetic field of the magnetar with negligible plasma effects.  It is now clear that our work focuses on the vacuum sector.

On the one hand, the ALP-to-photon conversion probability $P_{a\to\gamma}$ for pure ALP initial state in the limit of dipolar magnetic field depends on six ALP and magnetar parameters corresponding to the ALP energy $\omega$, the ALP mass $m_a$, the ALP-photon coupling constant $g$, the magnetar's radius $r_0$, the (dimensionless) magnetar magnetic field at the surface $b_0=B_0/B_c$ and the angle between the direction of propagation and the magnetic field $\theta$.  On the other hand, the normalized surface-subtracted Stokes parameter $R(\chi_0)\equiv R(\chi_0,\infty)$ depends also on the mixed initial state through $\chi_0$.

\begin{figure}[!t]
\centering
\resizebox{15cm}{!}{
\includegraphics{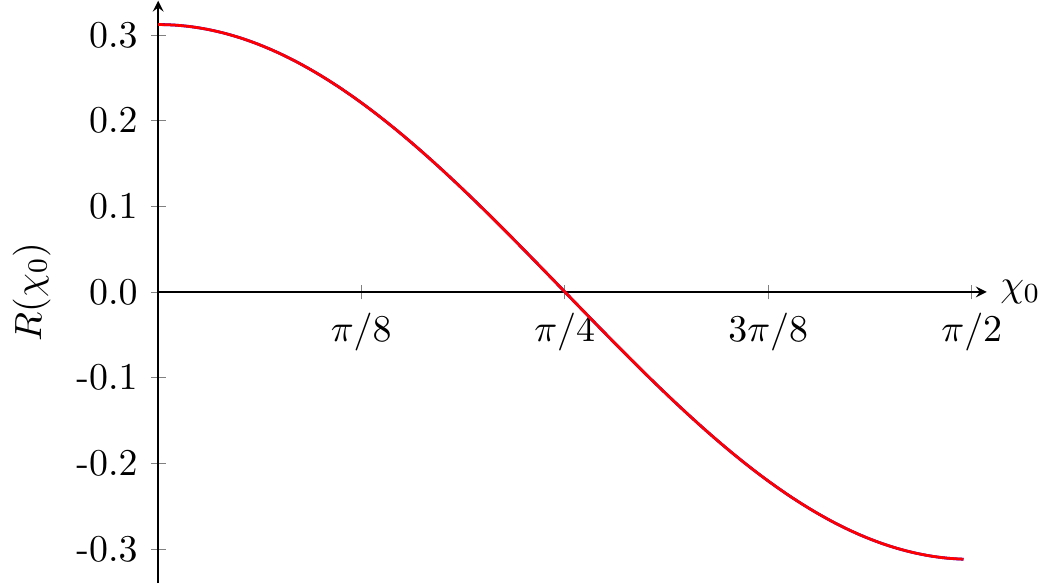}
\hspace{2cm}
\includegraphics{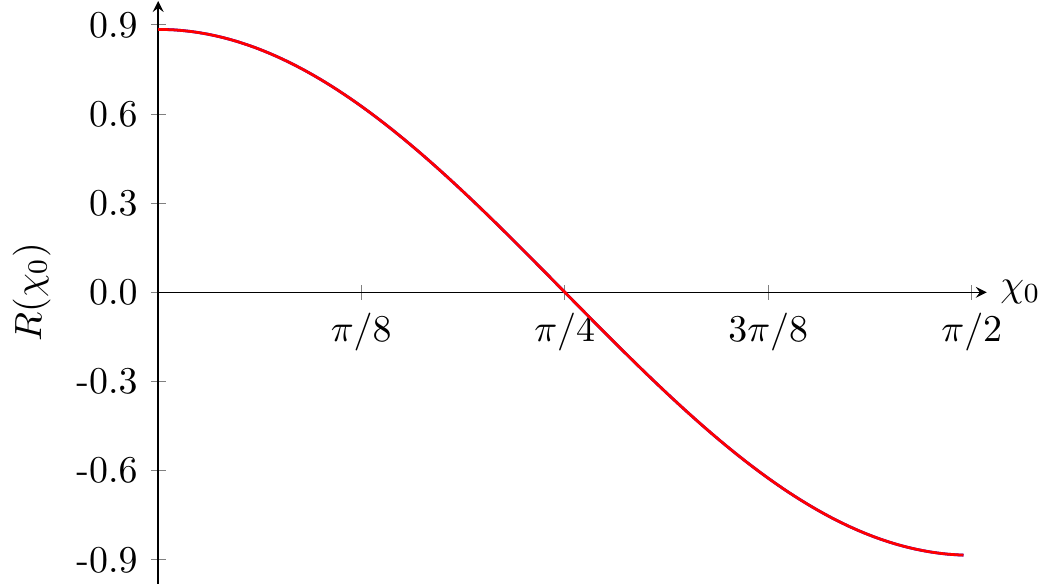}
}
\caption{Normalized surface-subtracted Stokes parameter $R$ as a function of $\chi_0$.  The blue lines correspond to $R(\chi_0)$ obtained from averaging over $\Delta\phi_0$ as dictated in \eqref{EqnR} while the red lines correspond to $P_{a\to\gamma}\cos(2\chi_0)$ following \eqref{EqnRSoln}.  The ALP and magnetar parameters are $\omega=1\,\text{keV}$, $m_a=10^{-9}\,\text{keV}$, $g/e=10^{-15}\,\text{keV}^{-1}$, $r_0=9\,\text{km}$, $B_0=10^{14}\,\text{G}$ and $\theta=\pi/4$ for the left panel while they are $\omega=10\,\text{keV}$, $m_a=10^{-8}\,\text{keV}$, $g/e=10^{-14}\,\text{keV}^{-1}$, $r_0=11\,\text{km}$, $B_0=10^{13}\,\text{G}$ and $\theta=-\pi/8$ for the right panel.}
\label{FigR}
\end{figure}

Fig.~\ref{FigR} shows the normalized surface-subtracted Stokes parameter \eqref{EqnR} for two points in the ALP and magnetar parameter space in function of $\chi_0$.  A comparison with $P_{a\to\gamma}\cos(2\chi_0)$ is also shown, confirming the validity of \eqref{EqnRSoln} (they are undistinguishable).

\begin{figure}[!t]
\centering
\resizebox{15cm}{!}{
\includegraphics{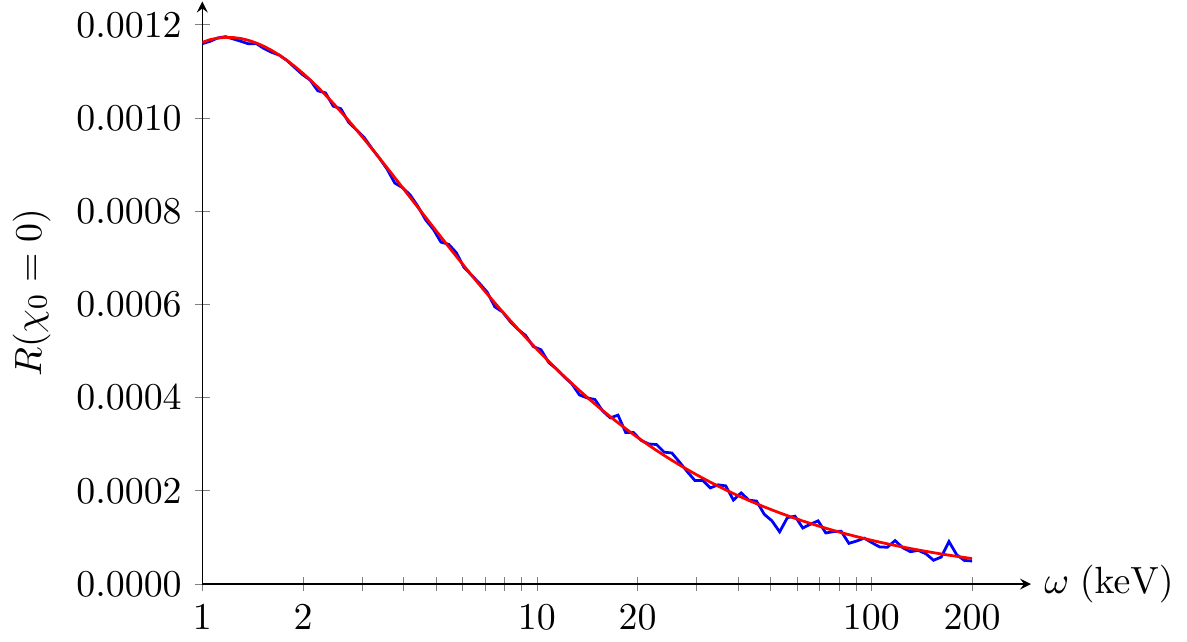}
\hspace{2cm}
\includegraphics{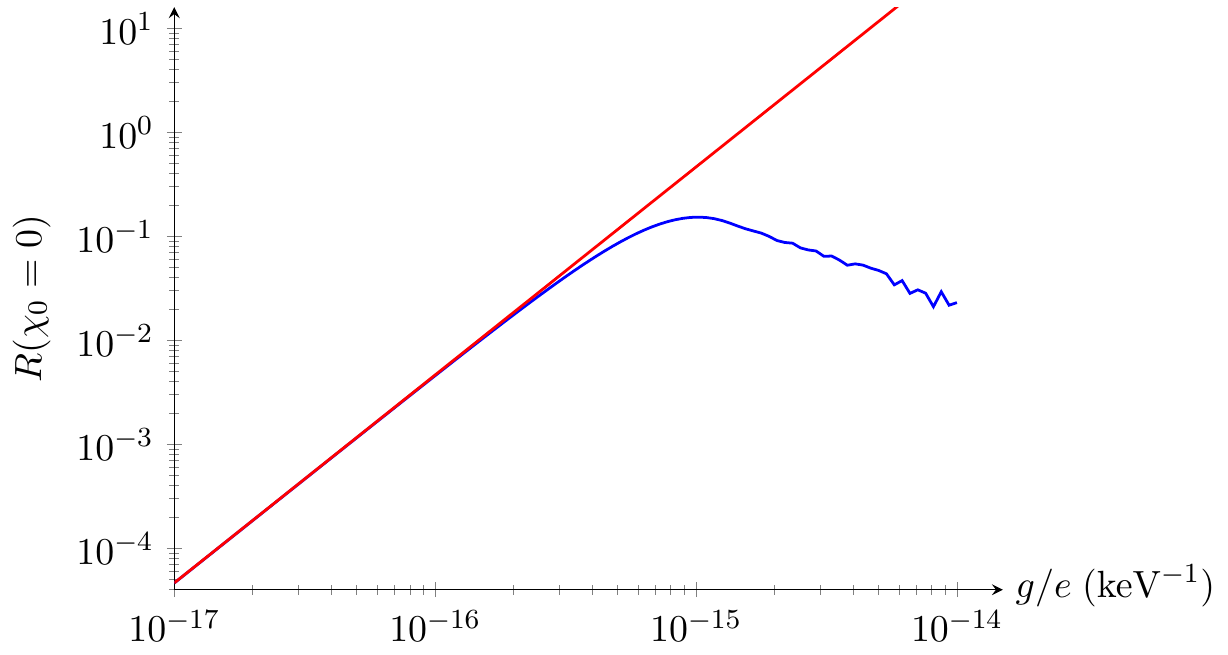}
}
\caption{Normalized surface-subtracted Stokes parameter $R$ for pure ALP initial state as a function of $\omega$ and $g$.  The blue lines are obtained from the evolution equations \eqref{EqnEE} while the red lines correspond to the approximation \eqref{EqnRapprox}.  The benchmark values of the ALP-photon coupling and magnetar parameters are $m_a=10^{-8}\,\text{keV}$, $r_0=10\,\text{km}$, $B_0=20\times10^{14}\,\text{G}$ and $\theta=\pi/2$.  For the left panel $g/e=5\times10^{-17}\,\text{keV}^{-1}$, while for the right panel $\omega=1\,\text{keV}$.}
\label{FigBP}
\end{figure}

Focusing now on the benchmark points $\omega=1$ and $100\,\text{keV}$, $m_a=10^{-8}\,\text{keV}$, $g/e=5\times10^{-17}\,\text{keV}^{-1}$, $r_0=10\,\text{km}$, $B_0=20\times10^{14}\,\text{G}$ and $\theta=\pi/2$, which lie in the soft and hard X-ray range with appropriate magnetar parameters for SGR 1806-20 and are not excluded by CAST, different dependences of $R(\chi_0=0)$ around the benchmark points are shown in Fig.~\ref{FigBP}.

As explained in \cite{Fortin:2018ehg}, the fact that the conversion probability peaks in the X-ray range (as shown in the first panel of Fig.~\ref{FigBP}) has interesting consequences since the ALP nucleon-nucleon bremsstrahlung emission spectrum peaks in the hard X-ray range for our benchmark model (for a degenerate medium relevant to magnetars).  The second panel shows $R$ as a function of the ALP-photon coupling constant $g$.  It is shown that the conversion probability drops dramatically for smaller ALP-photon coupling constant, as expected from physical arguments.  Indeed, for small ALP-photon coupling constant $g$, there is effectively no oscillations and the conversion probability decreases.

Furthermore, contrary to some of the benchmark points used in \cite{Fortin:2018ehg}, the conversion probability for the CAST-friendly benchmark points used here can be obtained by analogy to time-dependent perturbation theory in quantum mechanics, leading to the approximation \cite{Raffelt:1987im}
\eqna{
P_{a\to\gamma}(x)&=\left|\int_1^xdx'\,\Delta_M(x')r_0\,\text{exp}\left\{i\int_1^{x'}dx''\,[\Delta_a-\Delta_\parallel(x'')]r_0\right\}\right|^2\\
&=(\Delta_{M0}r_0)^2\left|\int_1^xdx'\,\frac{1}{x'^3}\,\text{exp}\left[i\Delta_ar_0\left(x'-\frac{x_{a\to\gamma}^6}{5x'^5}\right)\right]\right|^2.
}[EqnPapprox]
These equations are correct as long as the numerical value for $g$ is small enough for the approximation to make sense.  The second equality, written in terms of the dimensionless conversion radius \cite{Fortin:2018ehg}
\eqn{x_{a\to\gamma}=\frac{r_{a\to\gamma}}{r_0}=\left(\frac{7\alpha}{45\pi}\right)^{1/6}\left(\frac{\omega}{m_a}\frac{B_0}{B_c}|\sin\theta|\right)^{1/3},}[EqnConvRadius]
is valid in the large conversion radius limit since in that limit $\hat{q}_\parallel\to1$ and the integral in the exponential can be trivially computed.  Hence, from \eqref{EqnPapprox} in the large conversion radius limit the normalized surface-subtracted Stokes parameter $R(\chi_0)$ at infinity is given by
\eqn{R(\chi_0)=\left(\frac{\Delta_{M0}r_0^3}{r_{a\to\gamma}^2}\right)^2\cos(2\chi_0)\left|\int_\frac{r_0}{r_{a\to\gamma}}^\infty dt\,\frac{1}{t^3}\,\text{exp}\left[i\Delta_ar_{a\to\gamma}\left(t-\frac{1}{5t^5}\right)\right]\right|^2,}[EqnRapprox]
where the norm of the integral in \eqref{EqnRapprox} is a number of order one for our benchmark points.

Since the integral in \eqref{EqnRapprox} is valid in the large conversion radius limit, it is possible to approximate it by integrating from the origin to infinity.  Indeed, the integral is negligible in the interval $[0,r_0/r_{a\to\gamma}]$ due to the $1/(5t^5)$ term.  The integral in \eqref{EqnRapprox} can therefore be seen as a function of the product $\Delta_ar_{a\to\gamma}$ only.  It can be evaluated approximatively in the large and small $|\Delta_ar_{a\to\gamma}|$ regimes using the steepest descent method or a change of variables respectively, leading to
\eqn{R(\chi_0)=\left(\frac{\Delta_{M0}r_0^3}{r_{a\to\gamma}^2}\right)^2\cos(2\chi_0)\times\begin{cases}\frac{\pi}{3|\Delta_ar_{a\to\gamma}|}e^{\frac{6\Delta_ar_{a\to\gamma}}{5}}&|\Delta_ar_{a\to\gamma}|\gtrsim0.45\\\frac{\Gamma\left(\frac{2}{5}\right)^2}{5^\frac{6}{5}|\Delta_ar_{a\to\gamma}|^\frac{4}{5}}&|\Delta_ar_{a\to\gamma}|\lesssim0.45\end{cases}.}[EqnRapproxanal]

Fig.~\ref{FigBP} compares the normalized surface-subtracted Stokes parameter $R$ for pure ALP initial state obtained from the evolution equations \eqref{EqnEE} (blue) and the approximation \eqref{EqnRapprox} (red), showing that they agree for small $g$, including our benchmark value $g/e=5\times10^{-17}\,\text{keV}^{-1}$.  Moreover, for our benchmark point relevant to SGR 1806-20, the analytic solution with $|\Delta_ar_{a\to\gamma}|\gtrsim0.45$ is valid in the soft X-ray regime (\textit{i.e.} from its lower boundary at $\omega\sim0.1\,\text{keV}$ to $\omega\lesssim3.89\,\text{keV}$) while the analytic solution with $|\Delta_ar_{a\to\gamma}|\lesssim0.45$ is valid in the hard X-ray regime (\textit{i.e.} from $\omega\gtrsim3.89\,\text{keV}$ to its upper boundary at $\omega\sim200\,\text{keV}$) and beyond.  Both analytic solutions \eqref{EqnRapproxanal} overestimate $R$ \eqref{EqnRapprox} in the region around $\omega=3.89\,\text{keV}$.


\section{Observational Outlook}\label{SObs}

We now turn to some comments on the observational outlook of our methods.

The field of X-ray polarimetry is currently very active, with several new instrument designs being proposed and detectors that have either been launched or are in the planning stage \cite{fabiani}.  The theoretical targets include understanding the emission from white dwarfs and neutron stars in binary systems, and the coupling of the plasma to the magnetic field in accreting X-ray pulsars \cite{gnedin,Marin:2017vdv}.

The targets most germane to our work are neutron stars and magnetars.  For these stellar objects, there are two sources of polarization that have been explored theoretically: the polarization due to non-linear QED effects \cite{gnedin2} and the polarization due to the anisotropic opacities of the surface plasma \cite{Lai:2003nd,Taverna:2015vpa}.  While polarization due to ALP-photon mixing has been studied, this has mainly been in the context of soft thermal emission \cite{Lai:2006af}.

The specific observational features of our work are summarized below.

\subsection{Astrophysical Polarization of Surface Radiation}

The polarization of X-rays at the surface of magnetars has been extensively studied in the astrophysics community.  Here, we briefly summarize this vast literature.

The photon energies that are relevant for us are far below the electron cyclotron frequency, given by $\omega_{ce}=m_e(B/B_c)$.  In this regime, the photons can be described in terms of two normal modes: the ordinary (O-mode) and the extraordinary (X-mode) which are respectively parallel and perpendicular to the plane containing the external magnetic field and the direction of propagation.  The X-mode opacity is generally suppressed compared to the O-mode opacity by a factor \cite{Lai:2003nd}
\eqn{\frac{I_\parallel(\chi_0,x=1)}{I_\perp(x=1)}\sim\left(\frac{\omega}{\omega_{ce}}\right)^2=\left(\frac{\omega}{m_e}\frac{B_c}{B_0}\right)^2\sim7.46\times10^{-7}\left(\frac{\omega}{1\,\text{keV}}\frac{10^{14}\,\text{G}}{B_0}\right)^2.}[ratioOX]
This implies that the radiation that escapes from the magnetar atmosphere is almost fully linearly polarized.  Since the ALP and X-mode intensities are independent and it is assumed that ALP production is significant compared to photon production, we can recast \eqref{ratioOX} into the requirement that $\chi_0\sim0$.

There is an additional subtlety to the issues discussed above, namely, the effect of vacuum birefringence.  This has been considered by many authors (\textit{e.g.} \cite{Lai:2006af}), and we summarize the main points.  Generally, the combined contributions of plasma and vacuum polarization to the dielectric tensor can cause a vacuum resonance, where the X-mode and O-mode can convert into each other.  The relevant evolution of the O-mode and X-mode is given by the following equation,
\eqn{i\frac{d}{dr}\left(\begin{array}{c}E_\parallel\\E_\perp\end{array}\right)=\frac{\omega}{2}\left(\begin{array}{ccc}2+\sigma_{11}&\sigma_{12}\\\sigma_{21}&2+\sigma_{22}\end{array}\right)\left(\begin{array}{c}E_\parallel\\E_\perp\end{array}\right).}
In the above, the $\sigma_{ii}$ denote components of the dielectric tensor of the plasma, neglecting damping terms and in the limit that the proton is massive.  Since the electron cyclotron frequency is much larger than the relevant photon energies, we can take
\eqn{\sigma_{12}=-\sigma_{21}=0,\qquad\qquad\sigma_{11}=\left(q_\parallel-\frac{\omega^2_{pe}}{\omega^2}\right)\sin^2{\theta},\qquad\qquad\sigma_{22}=-q_\perp\sin^2{\theta},}
where we have used the electron plasma frequency $\omega_{pe}^2=4\pi e^2n_e/m_e$ with $n_e=Y_e\rho/m_p$ the electron density, $Y_e$ the electron fraction, $\rho$ the density and $m_p$ the proton mass.

A vacuum resonance can occur when $\sigma_{11}=\sigma_{22}$.  This happens when the contributions due to the plasma and QED exactly cancel, that is
\eqn{\frac{\omega^2_{pe}}{\omega^2}=q_\parallel+q_\perp.}
The corresponding resonant density is given by
\eqn{\rho\sim0.964Y_e^{-1}\left(\frac{\omega}{1\,\text{keV}}\frac{B}{10^{14}\,\text{G}}\right)^2\,\text{g}\cdot\text{cm}^{-3}.}[densityneeded]
When the plasma density reaches the value required for resonance, the dominant polarization mode can change, since the O-mode and X-mode photons can convert into each other and the two modes have different opacities.

From \eqref{densityneeded}, it is clear that for the strong magnetic fields that we are interested in, the required density for resonant conversion between O-modes and X-modes is extremely high.  The emergent radiation is thus largely dominated by the X-mode.

Besides thermal soft X-ray radiation, magnetars also exhibit considerable emission in the hard X-ray regime.  In \cite{Wadiasingh:2017rcq}, resonant inverse Compton scattering of thermal photons by ultra-relativistic charges was considered as the dominant production mechanism of hard X-rays.  The resulting spectrum was found to be strongly polarized in the X-mode, in the range of energies between $50\,\text{keV}-1\,\text{MeV}$, for a broad selection of viewing angles, magnetic field strengths $B/B_c\sim10-100$ and electron Lorentz factors $\gamma_e\sim10-100$.

Similarly, \cite{Beloborodov:2012ug} considered the injection of relativistic particles into the magnetosphere, which fill the large magnetic loops and spawn $e^+e^-$ pairs by scattering with photons.  In the outer parts of the loop where the magnetic field is reduced, the scattered photons are not energetic enough to spawn $e^+e^-$ pairs and are instead radiated out, giving rise to the hard X-ray spectrum.  A detailed analysis of the resulting spectrum by the authors showed that it is strongly polarized in the X-mode.

We note that in the non-thermal regime, the extent of polarization in the X-mode is dependent on the photon energy, electron Lorentz factor, field loop altitude and azimuth.  For the thermal regime, the extent of polarization is likewise dependent on the full physics of the plasma.  Nevertheless, from an astrophysical standpoint, the expectation is that for magnetars, the emission should be mostly polarized along the X-mode.

For the benchmark scenarios shown here, we will thus take $\chi_0=0$ as our initial condition, that is, we will assume that the photons coming from thermal and non-thermal emissions are completely polarized along the X-mode, and that any parallel mode photons come from ALP-to-photon conversion.  The case of $\chi_0>0$ can be obtained by simple rescaling.

\subsection{Polarization Limiting Radius}

We have seen that astrophysical models predict dominant X-mode polarization near the magnetar.  The polarization vector adiabatically tracks the variation of the direction of the dipolar magnetic field.  Moreover, the strong magnetic field in the magnetosphere endows the parallel and perpendicular modes with different refractive indices, preventing the mixing of modes and enhancing the polarization.

The adiabatic tracking of the polarization vector continues up to the polarization radius.  For a bipolar magnetic field, the radius is given by \cite{Heyl:2018kah}
\eqn{r_{PL}=\left(\frac{\alpha}{45}\frac{\nu}{c}\right)^{1/5}\left(\frac{B_0}{B_c}r_0^3\sin\beta\right)^{2/5}\sim923.4\left(\frac{\omega}{1\,\text{keV}}\right)^{1/5}\left(\frac{B_0}{10^{14}\,\text{G}}\right)^{2/5}\left(\frac{r_0}{10\,\text{km}}\right)^{6/5}\,\text{km},}
where $\beta$ is the angle between the dipole axis and the line of sight.

The observed X-mode polarization will be perpendicular to the magnetic field direction at the polarization radius.  Since the radius is large, the polarization will not track the magnetic field structure near the magnetar, but rather, will be oriented perpendicularly to the direction of the magnetic axis.

We thus note that the effect of QED is to enhance the observed degree of polarization up to 70\% or even more in the X-mode.  Thus, a large amount of polarization can be an indicator of non-linear QED effects \cite{Heyl:2018kah}.

\subsection{Effect of ALPs on Polarization Pattern}

In the previous subsections, we have discussed the polarization pattern expected at the polarization radius, which can be observed at upcoming polarization experiments.  We now turn to the effect of ALPs on this polarization pattern.
\begin{figure}[!t]
\centering
\resizebox{15cm}{!}{
\includegraphics{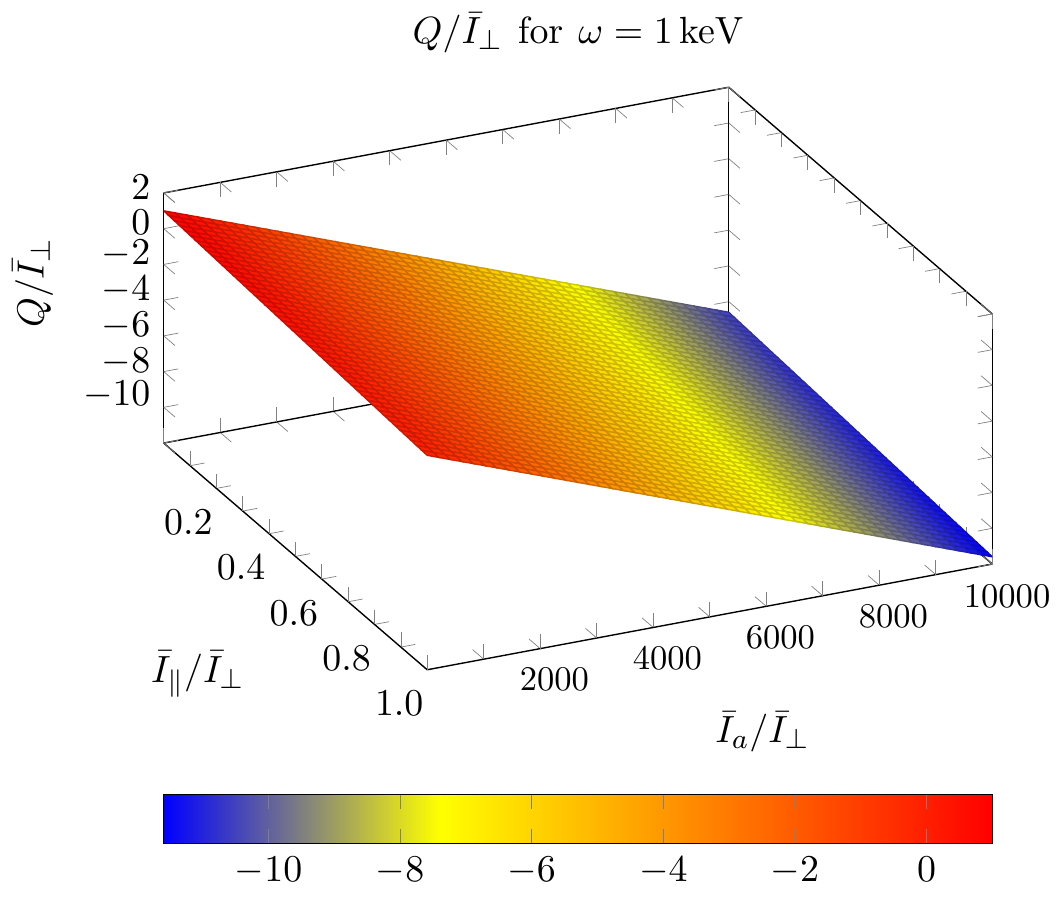}
\hspace{2cm}
\includegraphics{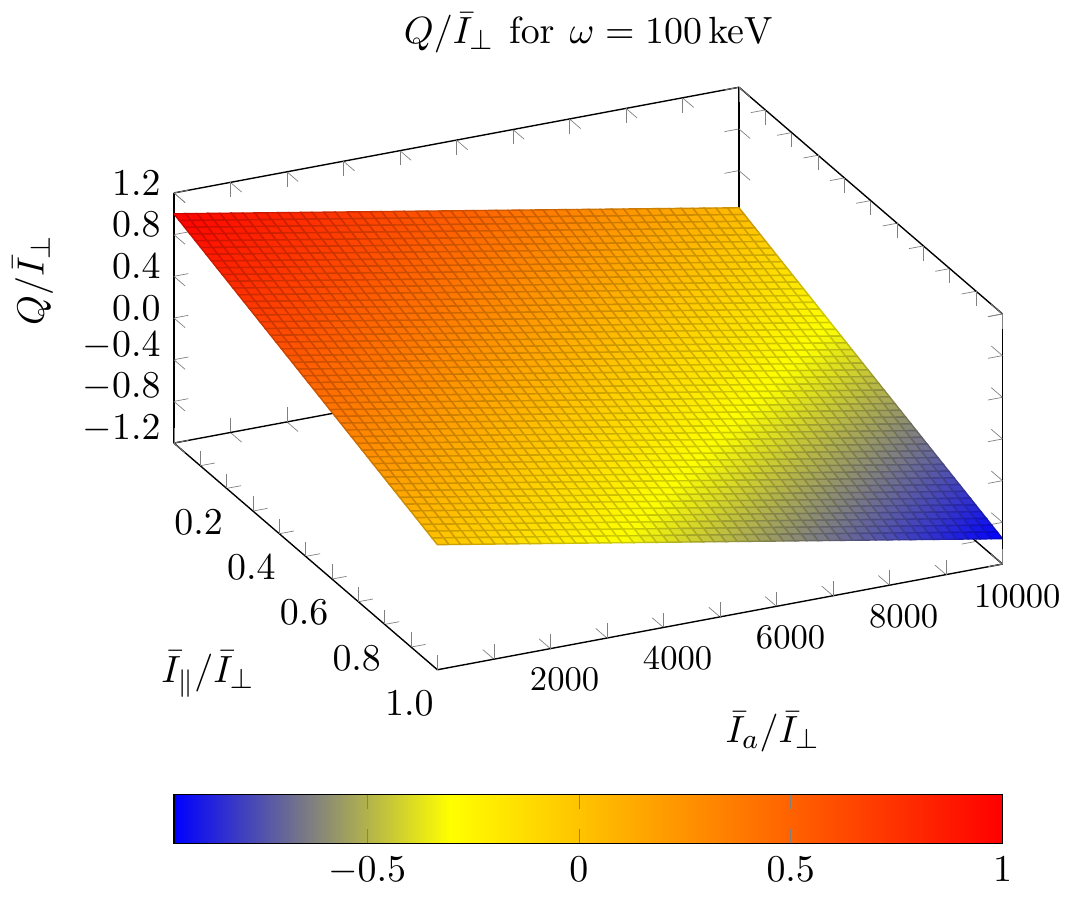}
}
\caption{(Normalized) Stokes parameter $Q/\bar{I}_\perp$ at infinity for the benchmark point $m_a=10^{-8}\,\text{keV}$, $g/e=5\times10^{-17}\,\text{keV}^{-1}$, $r_0=10\,\text{km}$, $B_0=20\times10^{14}\,\text{G}$ and $\theta=\pi/2$ in the plane $(\bar{I}_\parallel/\bar{I}_\perp,\bar{I}_a/\bar{I}_\perp)$ at the surface.  The left panel corresponds to ALP energy $\omega=1\,\text{keV}$ ($P_{a\to\gamma}\sim1.2\times10^{-3}$) while the right panel corresponds to $\omega=100\,\text{keV}$ ($P_{a\to\gamma}\sim9.4\times10^{-5}$).}
\label{FigStokesQ}
\end{figure}

ALPs produced inside the magnetar will escape into the magnetosphere with soft and hard X-ray energies.  Assuming that ALP production is large compared to the negligible O-mode photon production from astrophysics, the correct initial condition is $\chi_0=0$.  On the other hand, the relative magnitude of the ALP and X-mode intensities at the surface, $\bar{I}_a/\bar{I}_\perp$, depends on ALP production luminosity from the core.

The ALPs coming into the magnetosphere will eventually convert to O-mode photons.  The maximum conversion occurs around [see \eqref{EqnConvRadius}]
\eqna{
r_{a\to\gamma}&=\left(\frac{7\alpha}{45\pi}\right)^{1/6}\left(\frac{\omega}{m_a}\frac{B_0}{B_c}|\sin\theta|\right)^{1/3}r_0\\
&\sim1626.9\left(\frac{\omega}{1\,\text{keV}}\right)^{1/3}\left(\frac{B_0}{10^{14}\,\text{G}}\right)^{1/3}\left(\frac{10^{-8}\,\text{keV}}{m_a}\right)^{1/3}\left(\frac{r_0}{10\,\text{km}}\right)\,\text{km}.
}
We note that
\eqn{\frac{r_{PL}}{r_{a\to\gamma}}\sim0.57\left(\frac{1\,\text{keV}}{\omega}\right)^{2/15}\left(\frac{B_0}{10^{14}\,\text{G}}\right)^{1/15}\left(\frac{m_a}{10^{-8}\,\text{keV}}\right)^{1/3}\left(\frac{r_0}{10\,\text{km}}\right)^{1/5},}
and thus $r_{PL}\lesssim r_{a\to\gamma}$ for our benchmark points, implying that the effect of ALPs on the observed polarization pattern will be to add an O-mode intensity to the purely astrophysical X-mode intensity.

The previous effect can be demonstrated by the $Q$-parameter defined in \eqref{EqnStokesSoln}, which can be re-expressed as
\eqn{Q(x)=\bar{I}_\perp-\bar{I}_\parallel-(\bar{I}_a-\bar{I}_\parallel)P_{a\to\gamma}(x),}
where here $\bar{I}_\parallel\equiv\bar{I}_\parallel(1)$ and $\bar{I}_a\equiv\bar{I}_a(1)$ are surface intensities.  For example, the variation of $Q/\bar{I}_\perp$ at infinity in the plane $(\bar{I}_\parallel/\bar{I}_\perp,\bar{I}_a/\bar{I}_\perp)$ at the surface is shown in Fig.~\ref{FigStokesQ} for $\omega=1$ and $100\,\text{keV}$ with ALP and magnetar parameters $m_a=10^{-8}\,\text{keV}$, $g/e=5\times10^{-17}\,\text{keV}^{-1}$, $r_0=10\,\text{km}$, $B_0=20\times10^{14}\,\text{G}$ and $\theta=\pi/2$ relevant to SGR 1806-20.  The differences between the (normalized) Stokes parameter $Q/\bar{I}_\perp$ for $\omega=1$ and $100\,\text{keV}$ exhibited in Fig.~\ref{FigStokesQ} come solely from the different conversion probabilities.

\begin{figure}[!t]
\centering
\resizebox{15cm}{!}{
\includegraphics{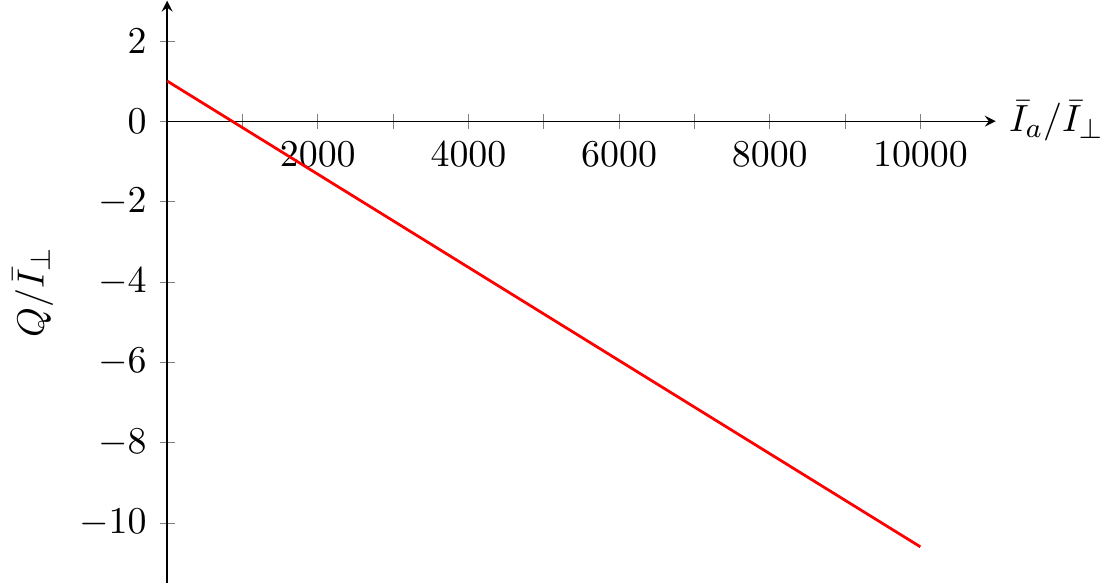}
\hspace{2cm}
\includegraphics{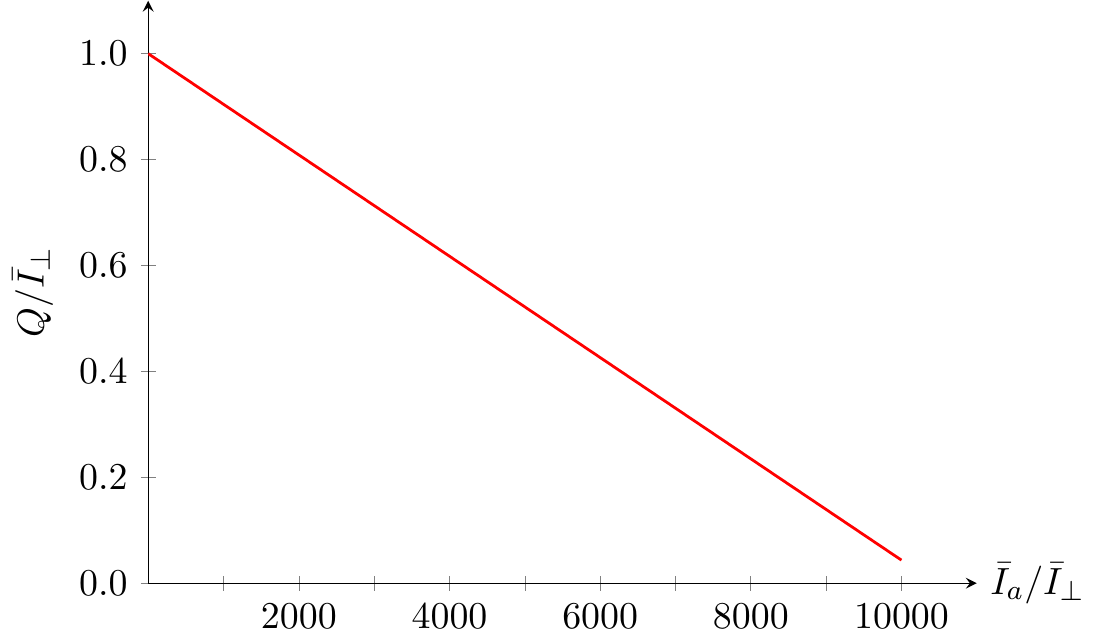}
}
\caption{(Normalized) Stokes parameter $Q/\bar{I}_\perp$ at infinity for the benchmark point $m_a=10^{-8}\,\text{keV}$, $g/e=5\times10^{-17}\,\text{keV}^{-1}$, $r_0=10\,\text{km}$, $B_0=20\times10^{14}\,\text{G}$ and $\theta=\pi/2$ in function of $\bar{I}_a/\bar{I}_\perp$ at the surface assuming $\bar{I}_\parallel/\bar{I}_\perp\sim0$ at the surface.  The left panel corresponds to ALP energy $\omega=1\,\text{keV}$ ($P_{a\to\gamma}\sim1.2\times10^{-3}$) while the right panel corresponds to $\omega=100\,\text{keV}$ ($P_{a\to\gamma}\sim9.4\times10^{-5}$).}
\label{FigQ}
\end{figure}
As we already argued above, since for magnetars with $B_0\gtrsim7\times10^{14}\,\text{G}$ radiation from the surface is dominated by the perpendicular mode \cite{Lai:2006af}, it is expected that $\bar{I}_\parallel/\bar{I}_\perp\sim0$ for SGR 1806-20.  On the one hand, for negligible magnetar ALP production, one has $\bar{I}_a/\bar{I}_\perp\sim0$ and the Stokes parameter is simply $Q\sim\bar{I}_\perp$.  On the other hand, for significant magnetar ALP production, $\bar{I}_a/\bar{I}_\perp$ can be quite large leading to negative Stokes parameter $Q$.

Indeed, it was argued in \cite{Fortin:2018ehg} that ALP production in magnetars (with a luminosity subdominant to neutrinos such that magnetar cooling is not disturbed\footnote{For comparison, this argument implies that
\eqn{g_{aN}\leq7.3\times10^{-16}\left(\frac{\rho}{\rho_0}\right)^{1/6}\sqrt{R_M}\left(\frac{T}{10^9\,\text{K}}\right)\,\text{keV}^{-1},}
where $\rho$ is the magnetar density, $\rho_0$ is the nuclear saturation density and $R_M\leq1$ is a suppression factor which turns on with the onset of proton and/or neutron superfluidity.  Hence the bound on $g_{aN}$ from magnetar cooling constraint is roughly of the same order of magnitude than the CAST constraint on $g$.}) could generate the necessary photon luminosity through ALP-to-photon conversion in the magnetosphere without violating the CAST bound.  For such a scenario, the ALP intensity at the surface $\bar{I}_a$ can be four or five orders of magnitude larger than the observed photon luminosity \cite{Beloborodov:2016mmx}, implying ratios as large as $\bar{I}_a/\bar{I}_\perp\sim\mathcal{O}(10^4)$, and negative Stokes parameters $Q\sim-10\bar{I}_\perp$ for $\omega=1\,\text{keV}$ or negligible Stokes parameter $Q\sim0$ for $\omega=100\,\text{keV}$ (see Fig.~\ref{FigQ}).  Observations of negative or vanishing Stokes parameter could then be seen as evidence of ALP production in magnetars.


\section{Conclusion}\label{SConclusion}

X-ray polarimetry is an emerging field in astronomy that is capable of probing both extended and compact objects.  Polarimetry is particularly suited to probe anisotropic couplings of the photon.  From the perspective of fundamental physics, one particularly important coupling that can be probed by polarimetry is the ALP-photon coupling.  Magnetars, with their extreme magnetic fields, are natural laboratories in this context.  

In this paper we investigated oscillations between ALPs and photons in the magnetosphere of a magnetar.  Since ALP and photon production mechanisms in the soft and hard X-ray regimes are independent, we remained agnostic with respect to the production mechanisms and studied the normalized surface-subtracted Stokes parameter averaged over the initial phase difference.

We then found that the normalized surface-subtracted Stokes parameter $R$, which is the normalized Stokes parameter $I$ at infinity minus the normalized Stokes parameter $I$ at the magnetar's surface, factorizes into the ALP-to-photon conversion probability for pure ALP initial state times a simple function of the ALP-parallel photon mixture of the initial state.  This highly non-trivial statement was then proved mathematically and checked numerically for several ALP and magnetar parameters.  The averaged intensities as well as the Stokes parameters at infinity were then obtained in terms of the parameters describing the initial state and the ALP-to-photon conversion probability.

Apart from the factorization property, $R$ has several important properties.  Firstly, it vanishes when ALPs do not exist or ALP-parallel photon oscillations do not occur.  Moreover, it is non-zero if ALP-parallel photon oscillations occur and the initial state is not an equal mixture of ALPs and parallel photons.  For benchmark values corresponding to SGR 1806-20, we exhibited the behavior of $R$ in the ALP mass and ALP-photon coupling plane for different ALP energies.  Our results show that the changes in the Stokes parameter $Q$ can be quite large for magnetars, implying large deviations from the no-ALP case.

Indeed, the main new feature of our analysis has been to study the changes in the polarization pattern in both the soft as well as the hard X-ray regimes of magnetars, due to ALPs.  In the absence of ALPs, astrophysical modeling of thermal and hard X-rays from magnetars predicts mainly X-mode polarization, for which the electric field is perpendicular to the plane containing the external magnetic field and the direction of propagation.  ALPs add to the astrophysical picture described above by producing O-mode photons, for which the electric field is parallel to the plane containing the external magnetic field and the direction of propagation.  We have computed the radius of conversion, where the probability of conversion becomes significant, and find that it is typically of the same order or larger than the polarization radius, implying an overall O-mode superposed on the X-mode coming purely from astrophysics.

There are several missions that are poised to explore these features.  Both the Imaging X-ray Polarimetry Explorer (IXPE) \cite{ixpe} and the Enhanced X-Ray Timing and Polarimetry Mission (eXTP) \cite{extp} missions will launch in the next few years, and look for signals in the $2-10\,\text{keV}$ range.  Among the missions that study hard X-rays, most are focused on solar flares and gamma-ray bursts.  However, X-Calibur and PolariS will be sensitive to signals in the $20-60\,\text{keV}$ and $10-80\,\text{keV}$ range from neutron stars, respectively \cite{xcalibur}.

Finally, since our general approach is applicable to other oscillation systems with the probability conservation property, one could investigate changes in polarizations for neutrino oscillation problems.


\ack{
JFF is supported by NSERC and FRQNT.  KS is supported by the U.~S.~Department of Energy grant DE-SC0009956.  We thank Henric Krawczynski and other members of the XPP Collaboration for useful discussions.  KS would like to thank the organizers of the Santa Fe Summer Workshop in Particle Physics for hospitality during the course of this work.
}


\setcounter{section}{0}
\renewcommand{\thesection}{\Alph{section}}

\section{Proof of the Factorization Property}\label{SApp}

To prove the factorization property, we focus on a generic two-state oscillation system given by
\eqn{i\frac{d\boldsymbol{a}(x)}{dx}=M(x)\boldsymbol{a}(x),}[EqnOsc]
where $M^T(x)=M(x)$ such that probability is conserved.\footnote{The case $M^\dagger(x)=M(x)$, which also leads to probability conservation, is left to the reader.}  With $a_1(x)=A\cos[\chi(x)]e^{-i\phi_1}$ and $a_2(x)=iA\sin[\chi(x)]e^{-i\phi_2}$, the evolution equations for \eqref{EqnOsc} are generalized to
\eqna{
\frac{d\chi(x)}{dx}&=-D(x)\cos[\Delta\phi(x)],\\
\frac{d\Delta\phi(x)}{dx}&=C(x)+2D(x)\cot[2\chi(x)]\sin[\Delta\phi(x)],
}[EqnOscEvol]
where $C(x)=M_{11}(x)-M_{22}(x)$ and $D(x)=M_{12}(x)=M_{21}(x)$.

We now introduce the quantity of interest
\eqna{
P(\chi_0,\Delta\phi_0,x)&=\frac{1}{2}\left\{1-\frac{\cos\left[\left.2\chi(x)\right|_{\chi(1)=\chi_0,\Delta\phi(1)=\Delta\phi_0}\right]}{\cos(2\chi_0)}\right\},\\
\bar{P}(\chi_0,x)&=\int_0^{2\pi}\frac{d\Delta\phi_0}{2\pi}\,P(\chi_0,\Delta\phi_0,x),
}[EqnP]
which is the integral appearing in the normalized surface-subtracted Stokes parameter \eqref{EqnR}.  From the evolution equations \eqref{EqnOscEvol}, it is easy to verify that $P(\chi_0,\Delta\phi_0,x)$ satisfies the following differential equation,
\eqna{
0&=\frac{d^3}{dx^3}P(\chi_0,\Delta\phi_0,x)-\ln[C(x)D(x)^2]'\frac{d^2}{dx^2}P(\chi_0,\Delta\phi_0,x)\\
&\phantom{=}\hspace{20pt}+\{C(x)^2+4D(x)^2+\ln[C(x)D(x)]'\ln[D(x)]'-\ln[D(x)]''\}\frac{d}{dx}P(\chi_0,\Delta\phi_0,x)\\
&\phantom{=}\hspace{20pt}+4D(x)^2\ln[D(x)/C(x)]'P(\chi_0,\Delta\phi_0,x)-2D(x)^2\ln[D(x)/C(x)]',
}[EqnPdiff]
with boundary conditions
\eqna{
P(\chi_0,\Delta\phi_0,x=1)&=0,\\
\left.\frac{d}{dx}P(\chi_0,\Delta\phi_0,x)\right|_{x=1}&=-D(1)\tan(2\chi_0)\cos(\Delta\phi_0),\\
\left.\frac{d^2}{dx^2}P(\chi_0,\Delta\phi_0,x)\right|_{x=1}&=2D(1)^2+C(1)D(1)\tan(2\chi_0)\sin(\Delta\phi_0)-D'(1)\tan(2\chi_0)\cos(\Delta\phi_0).
}[EqnPbound]
To simplify the notation in \eqref{EqnPdiff} and \eqref{EqnPbound}, we introduced primes to denote derivatives with respect to $x$.

On the one hand, the evolution equation \eqref{EqnPdiff} for $P(\chi_0,\Delta\phi_0,x)$ is $\chi_0$-independent while the boundary conditions \eqref{EqnPbound} for $P(\chi_0,\Delta\phi_0,x)$ are not.  Moreover the boundary conditions \eqref{EqnPbound} for $P(\chi_0,\Delta\phi_0,x)$ are also $\Delta\phi_0$-dependent, as expected.  On the other hand, with the average over $\Delta\phi_0$, both the evolution equation \eqref{EqnPdiff} and the boundary conditions \eqref{EqnPbound} for $\bar{P}(\chi_0,x)$ are $\chi_0$-independent, and thus we conclude that $\bar{P}(\chi_0,x)\equiv\bar{P}(x)$.  In particular, $\bar{P}(x)$ satisfies the same differential equation with the same boundary conditions than the conversion probability.  This demonstrates that $\bar{P}(\chi_0,x)\equiv\bar{P}(x)$ is $\chi_0$-independent and corresponds to the conversion probability $\bar{P}(x)=P_{a_1\to a_2}(x)=\sin^2[\left.\chi(x)\right|_{\chi(1)=0,\Delta\phi(1)=0}]$, proving the factorization property of the normalized surface-subtracted Stokes parameter \eqref{EqnRSoln}.

It is interesting to note that if the average over $\Delta\phi_0$ is not performed, the differential equation for the non-averaged quantity \eqref{EqnP} is the same than \eqref{EqnPdiff} but the boundary conditions \eqref{EqnPbound} are modified such that the first and second derivatives at $x=1$ of the non-averaged quantity depend on $\chi_0$ (and $\Delta\phi_0$).  The average is thus necessary for the factorization property to hold.  Another important point to mention is that the choice of the integrand in \eqref{EqnP} is crucial since different choices do not necessarily lead to the factorization property even if they evaluate to the original integrand $P(\chi_0,\Delta\phi_0,x)$ at $\chi_0=0$.  Moreover, it might be curious at first to see that $P(\chi_0,\Delta\phi_0,x)$ and $\bar{P}(\chi_0,x)$ satisfy a third-order differential equation instead of a second-order differential equation.  The reason why this occurs is most clearly understood from the non-averaged quantity, since without the average on $\Delta\phi_0$, one needs a third-order differential equation for the non-averaged quantity to have two genuine boundary conditions, the boundary condition of the non-averaged quantity being zero for any $\chi_0$ and $\Delta\phi_0$.

Finally, we verified numerically that the differential equation \eqref{EqnPdiff} and the boundary conditions \eqref{EqnPbound} lead to the same value of the conversion probability than the evolution equations \eqref{EqnEE}.


\bibliography{Stokes}

\end{document}